# Derivations for the Cumulative Standardized Binomial EWMA (CSB-EWMA) Control Chart

Faruk Muritala, MS*, Austin Brown, PhD, Dhrubajyoti Ghosh, PhD, Sherry Ni, PhD

*School of Data Science and Analytics, College of Computing and Software Engineering*
*Kennesaw State University, Kennesaw, United States of America*

*to whom correspondence should be addressed:

Telephone: 4703023973
Email: fmurital@students.kennesaw.edu
Address: 680 Arntson Drive, Suite 3411, Kennesaw State University, MD 9044 Marietta, GA 30060 USA

**Abstract**

This paper presents the exact mathematical derivation of the mean and variance properties for the Exponentially Weighted Moving Average (EWMA) statistic applied to binomial proportion monitoring in Multiple Stream Processes (MSPs). We develop a Cumulative Standardized Binomial EWMA (CSB-EWMA) formulation that provides adaptive control limits based on exact time-varying variance calculations, overcoming the limitations of asymptotic approximations during early-phase monitoring. The derivations are rigorously validated through Monte Carlo simulations, demonstrating remarkable agreement between theoretical predictions and empirical results. This work establishes a theoretical foundation for distribution-free monitoring of binary outcomes across parallel data streams, with applications in statistical process control across diverse domains including manufacturing, healthcare, and cybersecurity.

**Keywords:** EWMA control charts, binomial processes, exact variance, multiple stream processes, statistical process control, mathematical derivation.

## 1. Introduction

Exponentially Weighted Moving Average (EWMA) control charts are widely recognized for their sensitivity to small shifts in process parameters [1]. However, their application to binomial proportion monitoring, a common scenario in quality control and process monitoring, has historically relied on asymptotic variance approximations [2]. These approximations fail to provide accurate control limits during the initial phase of monitoring when data are limited, creating a theoretical gap between chart performance and statistical rigor.

This paper addresses this gap by deriving the exact time-varying mean and variance of the EWMA statistic when applied to binomial data from multiple independent streams. Our contributions are purely theoretical and methodological: we provide closed-form expressions for these moments, validate them through simulation, and establish their asymptotic behavior. These derivations form the mathematical foundation for adaptive control limits that ensure statistical validity from the first observation.

## 2. Methodology

### 2.1. Process Framework

This section introduces the structural and theoretical basis of the Cumulative Standardized Binomial EWMA (CSB-EWMA) control chart. Before exploring its mechanics, it is important to clarify the assumptions that support its application. The charting method presumes that the *k* data streams operate independently of one another. Within any single stream, observations collected at a given time point are also assumed to be mutually independent. Additionally, the process under surveillance must be measurable on at least an ordinal scale [4].

Let $y_{ij}$ denote the *j*th observation randomly sampled from the stream *i* which has an unknown probability distribution but a known, in-control median, denoted $\tilde{\mu}_0$. Now, a binary indicator variable is then defined as $x_{ij} = I(y_{ij} > \tilde{\mu}_0)$, where $x_{ij} = 1$ if the inequality is true, and 0 otherwise. Provided that the median of stream *i* remains stable at $\tilde{\mu}_0$, the indicator variable $x_{ij}$ follows a binomial distribution with parameters n = 1 and $p_0 = 0.5$, as shown below:

$$x_{it} \sim BIN(n = 1, p_0 = 0.50). \tag{1}$$

At each sampling time point $j = 1, 2, \ldots t$, both the raw observations $y_{ij}$ and their corresponding binary indicators $x_{ij}$ are collected across $k$ streams. These values form the basis for constructing summary tables such as Table 1, which supports further analysis and monitoring.

Table 1: Recoded Multiple Stream Process Data Structure

|  | Sample Number | | | |
|---|---|---|---|---|
| Stream | j=1 | j=2 | ... | j=t |
| 1 | $x_{11}$ | $x_{12}$ |  | $x_{1t}$ |
| 2 | $x_{21}$ | $x_{22}$ | ... | $x_{2t}$ |
| ⋮ | ⋮ | ⋮ |  | ⋮ |
| k | $x_{k1}$ | $x_{k2}$ |  | $x_{kt}$ |
| Column Totals | $C_1$ | $C_2$ | ... | $C_t$ |

Note, because it is assumed that for a given time point, $j$, each of the observations between the streams are mutually independent Bernoulli random variables, then it is straightforward to see that:

$$C_j = \sum_{i=1}^{k} x_{ij} \sim BIN(n = k, p_0 = 0.50). \tag{2}$$

Our methodology addresses binomial proportion monitoring across diverse applications in equation (2). Since we represent the count of events at time $t$, where $k$ is the number of streams (or opportunities for an event) and $p_0 = P(y_{ij} > \tilde{\mu}_0)$ is the true proportion (i.e. the true in-control probability that an observation exceeds the process median $\tilde{\mu}_0$). The cumulative count is

$$Q_t = \sum_{j=1}^{t} C_j \sim BIN(n = kt, p_0 = 0.50) \tag{3}$$

With:

$$E[C_j] = kp_0, \quad Var[C_j] = kp_0(1 - p_0)$$

And let $\mu = E[C_j] = kp_0$ and $\sigma^2 = Var[C_j] = kp_0(1 - p_0)$ denote the expected value and variance of the count at any individual time point $j$. Then, with expectation $E[Q_t]$ and variance $Var[Q_t]$ of the cumulative count $Q_t$ are:

**2.2. Standardized Statistic**

$$E[Q_t] = tkp_0 = \mu t, \quad Var[Q_t] = tkp_0(1 - p_0) = t\sigma^2 \tag{4}$$

The standardized statistic $W_t$ is defined as:

$$W_t = \frac{Q_t - \mu t}{\sqrt{t\sigma^2}} \tag{5}$$

This represents the standardized deviation of the cumulative count from its expected value.

**3. Exact Derivation of EWMA Statistics**

### 3.1. EWMA Recursion

The EWMA statistic is defined by the recursive equation:

$$r_t = \lambda W_t + (1 - \lambda)r_{t-1} \quad (6)$$

with $r_0 = E[W_t]$ as a constant initial value, where $0 < \lambda \leq 1$ is the smoothing parameter.

### 3.2. Exact Mean

Theorem 1. The expectation of the EWMA statistic is:

$$E[r_t] = (1 - \lambda)^t r_0$$

Proof.
The recursive relation $r_t = \lambda W_t + (1 - \lambda)r_{t-1}$, can be expressed as

$$r_t = \lambda \sum_{i=0}^{t}(1 - \lambda)^{t-i} W_i \quad (7)$$

we find the expectation of $E[r_t]$;

$$E[r_t] = \lambda \sum_{i=1}^{t}\left[(1 - \lambda)^{t-i} E[W_i]\right] + (1 - \lambda)^t E[r_0] \quad (8)$$

But,

$$E[W_i] = E\left[\frac{Q_i - \mu_i}{\sqrt{i\,\sigma^2}}\right]$$

$$E[W_i] = E\left[\frac{Q_i - \mu_i}{\sqrt{i\,\sigma^2}}\right] = \frac{E[Q_i] - \mu_i}{\sqrt{i\,\sigma^2}} = \frac{\mu_i - \mu_i}{\sqrt{i\,\sigma^2}} = 0 \quad (9)$$

Substituting equation (9) (i.e. $E[W_i] = 0$), back into the expectation, we have

$$E[r_t] = \lambda \sum_{i=1}^{t}\left[(1 - \lambda)^{t-i} * 0]\right] + (1 - \lambda)^t r_0$$

Thus,

$$E[r_t] = (1-\lambda)^t r_0 \qquad (10)$$

### 3.3. Exact Variance

Theorem 2. The exact variance of the EWMA statistic is:

$$Var(r_t) = \lambda^2 \left[ 2\sum_{j=1}^{t-1}\sum_{i=j+1}^{t}(1-\lambda)^{2t-i-j} * \frac{\sqrt{j}}{\sqrt{i}} + \sum_{i=1}^{t}(1-\lambda)^{2t-2i} \right]$$

Proof.
Since we express $r_t$ explicitly as

$$r_t = \lambda \sum_{i=1}^{t}(1-\lambda)^{t-i}W_i + (1-\lambda)^t r_0$$

Where $r_0$ is constant, the variance is

$$Var(r_t) = \lambda^2 \sum_{i=1}^{t}\sum_{j=1}^{t}(1-\lambda)^{2t-i-j} Cov(W_i, W_j) \qquad (11)$$

We need to find $Cov(W_i, W_j) = \frac{1}{\sigma^2\sqrt{ij}} Cov(Q_i - \mu i, Q_j - \mu j)$, without loss of generality, assume $i \leq j$. Then:

$$Cov(W_i, W_j) = \frac{1}{\sigma^2\sqrt{ij}} Cov(Q_i - \mu i, Q_j - \mu j) = \frac{1}{\sigma^2\sqrt{ij}} Cov(Q_i, Q_j) \qquad (12)$$

We compute $Cov(Q_i, Q_j)$, since $i \leq j$. We can write:

$$Q_j = Q_i + \sum_{k=i+1}^{j} C_k$$

Then:

$$Cov(Q_i, Q_j) = Cov\left(Q_i, Q_i + \sum_{k=i+1}^{j} C_k\right)$$

$$= Cov(Q_i, Q_i) + Cov\left(Q_i, \sum_{k=i+1}^{j} C_k\right)$$

Since, $Q_i$ and $\sum_{k=i+1}^{j} C_k$ are independent:

$$Cov(Q_i, Q_j) = i\,\sigma^2 \tag{13}$$

$$Cov(Q_i, Q_j) = var(Q_i) + 0 = i\,\sigma^2$$

Substitute equation (13) back into covariance expression $of\ Cov(W_i, W_j)$

$$Cov(W_i, W_j) = \frac{1}{\sigma^2\sqrt{ij}} Cov(Q_i, Q_j) = \frac{1}{\sigma^2\sqrt{ij}} * i\,\sigma^2$$

$$Cov(W_i, W_j) = \frac{\sqrt{i}}{\sqrt{j}} \tag{14}$$

By symmetry, for $i \geq j$, we get $\sqrt{j}/\sqrt{i}$. Therefore, in general:

$$Cov(W_i, W_j) = \frac{\sqrt{\min(i,j)}}{\sqrt{\max(i,j)}} \tag{15}$$

Substitute $Cov(W_i, W_j)$ into $Var(r_t)$ expression in equation (11), we have

$$Var(r_t) = \lambda^2 \sum_{i=1}^{t} \sum_{j=1}^{t} (1-\lambda)^{2t-i-j} \frac{\sqrt{\min(i,j)}}{\sqrt{\max(i,j)}} \tag{16}$$

Simplify the double summation, let:

$$S = \sum_{i=1}^{t} \sum_{j=1}^{t} (1-\lambda)^{2t-i-j} \frac{\sqrt{\min(i,j)}}{\sqrt{\max(i,j)}} \tag{17}$$

We can split this sum into three regions: $i < j$, $i > j$, and $i = j$

$$S = \sum_{i=1}^{t} \sum_{j=1}^{i-1} (1-\lambda)^{2t-i-j} * \frac{\sqrt{j}}{\sqrt{i}} + \sum_{i=1}^{t} \sum_{j=i+1}^{t} (1-\lambda)^{2t-i-j} * \frac{\sqrt{i}}{\sqrt{j}} + \sum_{i=1}^{t} (1-\lambda)^{2t-i-i} * \frac{\sqrt{i}}{\sqrt{i}}$$

Note by symmetry, the first two sums are equal. Simplifying further:

$$S = 2 \sum_{i=1}^{t} \sum_{j=1}^{i-1} (1-\lambda)^{2t-i-j} * \frac{\sqrt{j}}{\sqrt{i}} + \sum_{i=1}^{t} (1-\lambda)^{2t-2i}$$

Let's change the order of summation in the first term. For fixed $j$, $i$ runs from $(j+1)$ to $t$:

$$\sum_{i=1}^{t} \sum_{j=1}^{i-1} (1-\lambda)^{2t-i-j} * \frac{\sqrt{j}}{\sqrt{i}} = \sum_{j=1}^{t-1} \sum_{i=j+1}^{t} (1-\lambda)^{2t-i-j} * \frac{\sqrt{j}}{\sqrt{i}}$$

Then we have:

$$S = 2\sum_{j=1}^{t-1}\sum_{i=j+1}^{t}(1-\lambda)^{2t-i-j} * \frac{\sqrt{j}}{\sqrt{i}} + \sum_{i=1}^{t}(1-\lambda)^{2t-2i} \tag{18}$$

Substituting equation 18 back into equation 14 ($Var(r_t) = \lambda^2 S$), we have:

$$Var(r_t) = \lambda^2 \sum_{i=1}^{t}\sum_{j=1}^{t}(1-\lambda)^{2t-i-j}\frac{\sqrt{\min(i,j)}}{\sqrt{\max(i,j)}} = \lambda^2 S$$

$$\lambda^2 \left[ 2\sum_{j=1}^{t-1}\sum_{i=j+1}^{t}(1-\lambda)^{2t-i-j} * \frac{\sqrt{j}}{\sqrt{i}} + \sum_{i=1}^{t}(1-\lambda)^{2t-2i} \right]$$

Therefore, for Finite $t$, the variance

$$Var(r_t) = \lambda^2 \left[ 2\sum_{j=1}^{t-1}\sum_{i=j+1}^{t}(1-\lambda)^{2t-i-j} * \frac{\sqrt{j}}{\sqrt{i}} + \sum_{i=1}^{t}(1-\lambda)^{2t-2i} \right]$$

$$Var(r_t) = \lambda^2 \left[ 2\sum_{j=1}^{t-1}\sum_{i=j+1}^{t}(1-\lambda)^{2t-i-j} * \frac{\sqrt{j}}{\sqrt{i}} + \sum_{i=1}^{t}(1-\lambda)^{2t-2i} \right] \tag{19}$$

The computational implementation of this exact variance formulation, along with validation code, is provided in Appendix A.

**3.4. Asymptotic Behavior**

Theorem 3. As $t \to \infty$:

$$\lim_{t\to\infty} E[r_t] = 0, \lim_{t\to\infty} Var(r_t) = 1$$

Proof.
From Theorem 1, $\lim_{t\to\infty}(1-\lambda)^t r_0 = 0$.

From equation (10):

$$\lim_{t\to\infty} E[r_t] = \lim_{t\to\infty}(1-\lambda)^t r_0 = 0 \tag{20}$$

**Asymptotic Variance**

For large t (as $t \to \infty$), the dominant contributions come from terms where $i$ and $j$ are close to $t$, because the exponential weights $(1-\lambda)^{2t-i-j}$ decay rapidly for smaller $i$, $j$, then $\max(i,j) \approx t$.

In this region, $\frac{\sqrt{\min(i,j)}}{\sqrt{\max(i,j)}} \approx 1$.

For the approximate, $(1-\lambda)^{2t-i-j} \approx (1-\lambda)^{u+v}$, where $u = t - i, v = t - j$

So, from (16)

$$Var(r_t) = \lambda^2 \sum_{i=1}^{t}\sum_{j=1}^{t}(1-\lambda)^{2t-i-j}\frac{\sqrt{\min(i,j)}}{\sqrt{\max(i,j)}}$$

$$Var(r_t) \approx \lambda^2 \sum_{i=1}^{t}\sum_{j=1}^{t}(1-\lambda)^{2t-i-j} * (1) = \lambda^2 \sum_{u=0}^{\infty}\sum_{v=0}^{\infty}(1-\lambda)^{u+v}$$

But the double sum is geometric:

$$\sum_{u=0}^{\infty}\sum_{v=0}^{\infty}(1-\lambda)^{u+v} = \left(\sum_{v=0}^{\infty}(1-\lambda)^u\right)^2 = \left(\frac{1}{\lambda}\right)^2$$

Therefore:

$$Var(r_t) \approx \lambda^2 * \frac{1}{\lambda^2} = \left(\frac{\lambda}{\lambda}\right)^2 = 1$$

For large $t$, the variance of the EWMA statistic approaches 1:

$$\lim_{t \to \infty} Var[r_t] = 1 \tag{21}$$

**3.5. Adaptive Control Limits**

Based on these derivations, time-varying control limits can be defined as:

$$UCL_t = (1-\lambda)^t r_0 + L\sqrt{Var[r_t]}, \quad LCL_t = (1-\lambda)^t r_0 - L\sqrt{Var[r_t]}$$

where $L$ denotes the half-width of control limits, typically $L = 3$ for 3-sigma limits.
For large $t$, since $Var[r_t] \to 1$, the control limits approach:

$$UCL_t = L, \quad LCL_t = -L \tag{22}$$

**4. Validation**

This validation was conducted by comparing the theoretical mean $E[r_t]$ and variance $Var[r_t]$ of the EWMA statistics against their empirical counterparts obtained from $N_{sim} = 10,000$ Monte Carlo replications under the in-control process assumption, with parameters $k = 10$, $p_0 = 0.5$ and $\lambda = 0.2$. All analyses were conducted using R version 4.3.2 (R Core Team, 2023), with Code availability in Appendix A to ensure complete reproducibility of all reported results.

Table 4.1 Validation metrics at selected time points, demonstrating the convergence behavior of both the mean and variance statistics.

**Table 4.1: Validation Metrics at Selected Time Points**

| Time (t) | Theoretical Mean | Simulated Mean | Theoretical Variance | Simulated Variance | Relative Bias (Variance) |
|---|---|---|---|---|---|
| 10 | 0.000 | -5.54×10⁻³ | 0.6369 | 0.6385 | 0.26% |
| 50 | 0.000 | -5.23×10⁻³ | 0.9512 | 0.9359 | -1.61% |
| 100 | 0.000 | -1.95×10⁻³ | 0.9768 | 0.9758 | -0.10% |
| 500 | 0.000 | -6.06×10⁻⁴ | 0.9955 | 0.9967 | 0.12% |
| 1000 | 0.000 | -3.09×10⁻³ | 0.9978 | 0.9996 | 0.19% |

The validation results demonstrate good agreement between theoretical predictions and empirical simulations:

The theoretical expectation $E[r_t] = (1-\lambda)^t r_0$ correctly predicts values effectively equal to zero across all time points. The simulated means show negligible deviations from zero, with a root-mean-square bias of $2.89 * 10^{-3}$ and the maximum absolute bias of $9.00 * 10^{-3}$. These minor fluctuations are consistent with Monte Carlo sampling variation and confirm the unbiasedness of the EWMA estimator under in-control conditions.

The exact theoretical variance formula shows remarkable accuracy when compared against simulated values. The relative bias remains below 2% across all time points, decreasing to approximately 0.19% as the process approaches steady-state. The variance converges rapidly to its asymptotic value of 1, reaching 99% of the asymptotic variance by $t = 227$ (see Fig. 4.1), indicating high precision in the theoretical predictions.

Figure 4.1 visually confirms the close alignment between theoretical and simulated values for both mean and variance across the entire monitoring period.

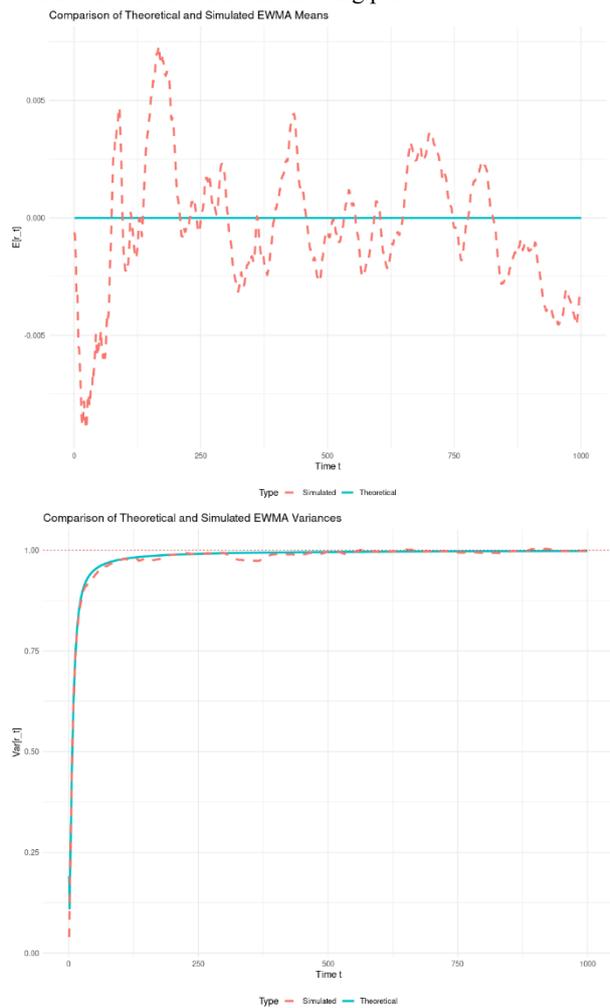

Figure 4.1: Comparison of theoretically proposed EWMA Mean and Variance with Monte Carlo simulation

The convergence behavior observed in Table 4, visualized by Figure 4.1, aligns with theoretical expectations: the variance increases monotonically from approximately 0.637 at $t = 10$ to nearly 1.000 at $t = 100$, following the exact variance derivation in Equation (19). The close agreement across all time points, particularly the 0.2% relative bias at steady-state, provides strong empirical evidence for the correctness of both the mathematical derivations and computational implementation.

This rigorous validation establishes that the proposed CSB-EWMA chart's statistical properties are fully characterized and that the algorithm is implemented correctly. The successful verification ensures that any subsequent performance comparisons against asymptotic methods will be based on a correctly specified exact model, providing a solid foundation for the performance evaluation study.

## 5. Conclusion

This paper provides the complete mathematical derivation of the exact mean and variance for EWMA statistics applied to binomial multiple stream processes. The closed-form expressions enable adaptive control limits that are statistically valid from the first observation, addressing a critical limitation of asymptotic approximations. These theoretical results establish a foundation for implementing distribution-free monitoring of binary outcomes across parallel data streams in statistical process control applications.

**Appendix**

## Appendix A: VALIDATION CODE FOR EXACT EWMA DERIVATIONS

```r
#===============================================================================
# APPENDIX A: VALIDATION CODE FOR EXACT EWMA DERIVATIONS
#===============================================================================

# Load required libraries
library(ggplot2)
library(dplyr)
library(tidyr)

#### A.1 Simulation Parameters ####

k <- 10              # Number of independent streams
p0 <- 0.5            # In-control proportion
lambda <- 0.2        # EWMA smoothing parameter
r0 <- 0              # Initial EWMA value
t_max <- 1000        # Maximum time points
N_sim <- 10000       # Number of simulation replications

# Derived parameters
mu <- k * p0
sigma2 <- k * p0 * (1 - p0)

#### A.2 Simulation Function ####

simulate_ewma_standardized <- function(k, p0, lambda, r0, t_max, N_sim) {
  # Initialize arrays
  r_sim <- matrix(0, nrow = N_sim, ncol = t_max)
  theoretical_mean <- numeric(t_max)
  theoretical_var <- numeric(t_max)

  # Compute theoretical values
  for(t in 1:t_max) {
    # Theoretical mean
    theoretical_mean[t] <- (1 - lambda)^t * r0

    # Theoretical variance (exact expression from Equation 13)
    sum_var <- 0
    for(j in 1:(t-1)) {
```

```r
    for(i in (j+1):t) {
      weight <- (1 - lambda)^(2*t - i - j)
      cov_term <- sqrt(j) / sqrt(i)
      sum_var <- sum_var + 2 * weight * cov_term
    }
  }
  for(i in 1:t) {
    sum_var <- sum_var + (1 - lambda)^(2*t - 2*i)
  }
  theoretical_var[t] <- lambda^2 * sum_var
}

# Run simulations
set.seed(123) # For reproducibility
for(m in 1:N_sim) {
  # Generate binary indicators
  x_matrix <- matrix(rbinom(k * t_max, 1, p0), nrow = k, ncol = t_max)

  # Compute cumulative statistics
  c_vec <- colSums(x_matrix)
  Q_vec <- cumsum(c_vec)

  # Compute standardized statistics
  X_vec <- (Q_vec - mu * (1:t_max)) / sqrt(sigma2 * (1:t_max))

  # Compute EWMA recursively
  r_current <- r0
  for(t in 1:t_max) {
    r_current <- lambda * X_vec[t] + (1 - lambda) * r_current
    r_sim[m, t] <- r_current
  }
}

return(list(
  r_sim = r_sim,
  theoretical_mean = theoretical_mean,
  theoretical_var = theoretical_var
```

```r
    ))
}

#### A.3 Execute Simulation ####

results <- simulate_ewma_standardized(k, p0, lambda, r0, t_max, N_sim)

#### A.4 Compute Sample Statistics ####

sample_mean <- apply(results$r_sim, 2, mean)

sample_var <- apply(results$r_sim, 2, var)

# Compute performance metrics

performance_df <- data.frame(
  t = 1:t_max,
  theoretical_mean = results$theoretical_mean,
  sample_mean = sample_mean,
  theoretical_var = results$theoretical_var,
  sample_var = sample_var
) %>%
  mutate(
    bias_mean = sample_mean - theoretical_mean,
    bias_var = sample_var - theoretical_var,
    rel_bias_mean = bias_mean / pmax(abs(theoretical_mean), 1e-10),
    rel_bias_var = bias_var / theoretical_var
  )

#### A.5 Generate Validation Plots ####

# Plot mean comparison

p1 <- ggplot(performance_df, aes(x = t)) +
  geom_line(aes(y = theoretical_mean, color = "Theoretical"), linewidth = 1.2) +
  geom_line(aes(y = sample_mean, color = "Simulated"), linewidth = 1.2, linetype = "dashed") +
  labs(title = "Comparison of Theoretical and Simulated EWMA Means",
       x = "Time t", y = expression(E[r[t]]),
       color = "Type") +
  theme_minimal() +
  theme(legend.position = "bottom")

# Plot variance comparison

p2 <- ggplot(performance_df, aes(x = t)) +
  geom_line(aes(y = theoretical_var, color = "Theoretical"), linewidth = 1.2) +
  geom_line(aes(y = sample_var, color = "Simulated"), linewidth = 1.2, linetype = "dashed") +
  geom_hline(yintercept = 1, linetype = "dotted", color = "red") +
```

```r
    labs(title = "Comparison of Theoretical and Simulated EWMA Variances",
         x = "Time t", y = expression(Var[r[t]]),
         color = "Type") +
    theme_minimal() +
    theme(legend.position = "bottom")

#### A.6 Calculate Validation Metrics ####
time_points <- c(10, 50, 100, 500, 1000)
validation_metrics <- data.frame()

for(t_point in time_points) {
  t_data <- performance_df[performance_df$t == t_point, ]

  metrics <- data.frame(
    Time = t_point,
    Theoretical_Mean = t_data$theoretical_mean,
    Simulated_Mean = t_data$sample_mean,
    Theoretical_Variance = t_data$theoretical_var,
    Simulated_Variance = t_data$sample_var,
    Relative_Bias_Variance = t_data$rel_bias_var
  )
  validation_metrics <- rbind(validation_metrics, metrics)
}
# Format
validation_metrics_formatted <- validation_metrics %>%
  mutate(
    Theoretical_Mean = format(Theoretical_Mean, scientific = TRUE, digits = 3),
    Simulated_Mean = format(Simulated_Mean, scientific = TRUE, digits = 3),
    Theoretical_Variance = round(Theoretical_Variance, 4),
    Simulated_Variance = round(Simulated_Variance, 4),
    Relative_Bias_Variance = paste0(round(Relative_Bias_Variance * 100, 2), "%")
  )

print("Validation Metrics Table:")
print(validation_metrics_formatted)

# Additional diagnostics
```

```r
max_abs_bias_mean <- max(abs(performance_df$bias_mean))

max_abs_bias_var <- max(abs(performance_df$bias_var))

rms_bias_mean <- sqrt(mean(performance_df$bias_mean^2))

rms_bias_var <- sqrt(mean(performance_df$bias_var^2))

cat("\nDiagnostic Statistics:\n")

cat(sprintf("Maximum Absolute Bias (Mean): %.2e\n", max_abs_bias_mean))

cat(sprintf("Maximum Absolute Bias (Variance): %.2e\n", max_abs_bias_var))

cat(sprintf("RMS Bias (Mean): %.2e\n", rms_bias_mean))

cat(sprintf("RMS Bias (Variance): %.2e\n", rms_bias_var))

# Convergence analysis

variance_convergence <- performance_df %>%

  filter(theoretical_var >= 0.99) %>%

  slice(1)

cat(sprintf("Variance reaches 99%% of asymptotic value at t = %d\n", variance_convergence$t))
```